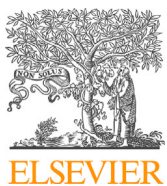
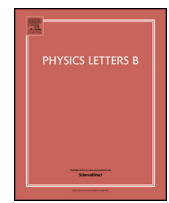

Letter

# Quantum-annealing-inspired algorithms for multijet clustering

Hideki Okawa [a],[*], Xian-Zhe Tao [b,c,d,e], Qing-Guo Zeng [b,c,d,e], Man-Hong Yung [b,c,d,e]

[a] *Institute of High Energy Physics, Chinese Academy of Sciences, Beijing, 100049, China*
[b] *Shenzhen Institute for Quantum Science and Engineering, Southern University of Science and Technology, Shenzhen, 518055, China*
[c] *International Quantum Academy, Shenzhen, 518048, China*
[d] *Guangdong Provincial Key Laboratory of Quantum Science and Engineering, Southern University of Science and Technology, Shenzhen, 518055, China*
[e] *Shenzhen Key Laboratory of Quantum Science and Engineering, Southern University of Science and Technology, Shenzhen, 518055, China*

A B S T R A C T

Jet clustering or reconstruction is a crucial component at high-energy colliders, a procedure to identify sprays of collimated particles originating from the fragmentation and hadronization of quarks and gluons. It is a complicated combinatorial optimization problem and requires intensive computing resources. In this study, we formulate jet reconstruction as a quadratic unconstrained binary optimization (QUBO) problem and introduce novel quantum-annealing-inspired algorithms for clustering multiple jets in electron-positron collision events. One of these quantum-annealing-inspired algorithms, ballistic simulated bifurcation, overcomes problems previously observed in multijet clustering with quantum-annealing approaches. We find that both the distance defined in the QUBO matrix and the prediction power of the QUBO solvers have crucial impacts on the multijet clustering performance. This study opens up a new approach to globally reconstructing multijet beyond dijet in one go, in contrast to the traditional iterative method.

## 1. Introduction

Jet clustering is a fundamental component at high-energy colliders that determines the kinematics of the underlying processes governed by quantum chromodynamics (QCD). Due to color confinement, quarks and gluons produced by the collisions or the decays of heavy particles initiate sprays of collimated particles originating from their fragmentation and hadronization. Jets serve as reliable proxies to determine the original parton kinematics.

Jet reconstruction is a complicated combinatorial optimization problem and requires intensive computing resources. It dates back to the proposal by Sternman and Weinberg [1], and various algorithms have been developed over decades since then, as reviewed in Refs. [2–6]. The majority of widely used jet clustering algorithms are implemented in FastJet [7] for both hadron and electron-positron colliders and have successfully been used in various experiments, including those at the Large Hadron Collider (LHC) [8–12].

As we face the unprecedented increase in luminosity at the High-Luminosity LHC (HL-LHC) [13] and future colliders under consideration, such as the Circular Electron Positron Collider (CEPC) [14–16], new approaches are being investigated actively to overcome this challenge. Applications of quantum computing and algorithms have recently attracted much attention and have been applied to track reconstruction, for example [17–25]. It also led to recent investigations on jet reconstruction using quantum annealing (QA) [26–29] and quantum gate machines [30–32].

This work demonstrates the potential of quantum-annealing-inspired algorithms (QAIAs) to pursue multijet clustering. In previous works using quantum annealing [26–28], reconstructing dijet has been pursued either with the thrust- or quantum-angle-based approach, with the latter providing higher performance than the former. However, there was a degradation in performance in multijet reconstruction [28], as multiple qubits are required to implement "one-hot" encoding and are error-prone [33]. By replacing the quantum angle with the $ee$-$k_t$ distance in the algorithm and using QAIAs, detailed in Section 2, we overcome the problem and maintain or even slightly improve performance from the traditional methods for the multijet reconstruction. Furthermore, as it is a "quantum-inspired" approach, our algorithms run on classical computers, and thus neither suffer from quantum hardware noise nor the limitations of the data size that they can handle. This study opens up a new approach to globally reconstructing multijet in one go, in contrast to the traditional iterative methods.

* Corresponding author.
  *E-mail address:* okawa@ihep.ac.cn (H. Okawa).

## 2. Methodology

Jet clustering can be regarded as a combinatorial optimization problem, formulated as a quadratic unconstrained binary optimization (QUBO) or Ising problem [34]. The difference between the two lies in using zero/one binaries for the former or the $\pm 1$ spins for the latter. The problem is designed so that the ground state of the QUBO/Ising model provides the correct answer. It is an NP (Nondeterministic Polynomial time) complete problem, and the solution candidates diverge exponentially with the problem size. QA machines by D-Wave based on the concept described in Ref. [35] and the coherent Ising machine (CIM) [36] are developed to solve such kinds of problems efficiently, for example. However, QA generally provides suboptimal results when handling largely connected graphs due to the limited connectivity of qubits and hardware noise [37–40]. Such a trend is consistently observed in the previous jet clustering studies [26,27]. This study introduces simulated bifurcation (SB), which overcomes those challenges.

### 2.1. Simulated bifurcation

The SB algorithm [41] emulates the quantum bifurcation machine (QbM) [42,43]. It solves combinatorial optimization problems through quantum adiabatic evolution of Kerr-nonlinear parametric oscillators, exhibiting bifurcation phenomena representing the two Ising spin states. SB can update all the Ising problem spins in parallel, allowing us to achieve computational acceleration. As stated above, solving the Ising problem is to find a spin configuration $\{x_i\}_{i=1}^N \in \{-1,1\}^N$ that minimizes the Hamiltonian $H(x_i)$ of the Ising model:

$$H(x_i) = \frac{1}{2}\sum_{ij}^N J_{ij} x_i x_j + \sum_i^N h_i x_i, \tag{1}$$

where $J_{ij}$ represents the spin-spin interactions and $h_i$ is the external field. In QbM, the Ising model is coupled to the Kerr-nonlinear parametric oscillators. According to the adiabatic evolution theory, if we set the initial state to the ground state of the system and the Hamiltonian changes gradually, the system will remain in the ground state throughout the evolution. Thus, in the end, we can obtain the ground state of the Ising problem. The corresponding classical analog, classical bifurcation machines (CbM) [42,43], are derived by approximating the expectation value of the annihilation operator with a complex amplitude $x_i + iy_i$. $x_i$ and $y_i$ are, respectively, the position and momentum of the $i$-th Kerr-nonlinear oscillator corresponding to the $i$-th spin. The original version of SB, adiabatic SB (aSB) [41], simplifies and improves CbM but is prone to errors originating from the continuous treatment of the spins in the differential equations. Two variants of SB are introduced to suppress such analog errors: ballistic SB (bSB) and discrete SB (dSB) [44]. The former introduces inelastic walls at $x_i = \pm 1$ as follows:

$$\dot{x}_i = \frac{\partial H_{\text{bSB}}}{\partial y_i} = a_0 y_i, \tag{2}$$

$$\dot{y}_i = -\frac{\partial H_{\text{bSB}}}{\partial x_i},$$

$$= -\left[a_0 - a(t)\right] x_i + c_0 \left(h_i + \sum_{j=1}^N J_{ij} x_j\right), \tag{3}$$

$$H_{\text{bSB}} = \frac{a_0}{2}\sum_{i=1}^N y_i^2 + V_{\text{bSB}}, \tag{4}$$

$$V_{\text{bSB}} = \begin{cases} \frac{a_0-a(t)}{2}\sum_{i=1}^N x_i^2 - c_0\left(\frac{1}{2}\sum_{i,j}^N J_{ij} x_i x_j + \sum_i h_i x_i\right), \\ \forall x_i,\ |x_i| \leq 1 \\ \infty,\ \text{otherwise} \end{cases} \tag{5}$$

where $a_0$ and $c_0$ are positive constants (the detuning frequency for the former and the coupling strength for the latter), $a(t)$ is a time-dependent pumping amplitude that monotonically increases from zero to $a_0$, $H_{\text{bSB}}$ is the Hamiltonian, and $V_{\text{bSB}}$ is the potential energy in bSB.

To further suppress the error from continuous relaxation of $x_i$, dSB discretizes $x_j$ to $\text{sgn}(x_j)$ in the mean-field term:

$$\dot{x}_i = \frac{\partial H_{\text{dSB}}}{\partial y_i} = a_0 y_i, \tag{6}$$

$$\dot{y}_i = \frac{\partial H_{\text{dSB}}}{\partial x_i} = -[a_0 - a(t)]x_i + c_0\left(\sum_{j=1}^N J_{ij}\text{sgn}(x_j) + h_i\right), \tag{7}$$

$$H_{\text{dSB}} = \frac{a_0}{2}\sum_{i=1}^N y_i^2 + V_{\text{dSB}}, \tag{8}$$

$$V_{\text{dSB}} = \begin{cases} \frac{a_0-a(t)}{2}\sum_{i=1}^N x_i^2 - c_0\left(\frac{1}{2}\sum_{i,j}^N J_{ij} x_i \text{sgn}(x_j) + \sum_j^N h_i x_i\right), \\ \forall x_i,\ |x_i| \leq 1 \\ \infty.\ \text{otherwise} \end{cases} \tag{9}$$

Finally, $\text{sgn}(x_i)$ gives the solution to the Ising problem. For all the SB variants, the symplectic Euler method is adopted for the numerical computation [45].

The SB algorithms used in this study are implemented in MindSpore Quantum [46–48]. A simulated annealing library, D-Wave Neal [49], is adopted as a benchmark for comparing to the SB algorithms.

### 2.2. Jet reconstruction as an Ising problem

As proposed in Refs. [26–28], jet reconstruction can be formulated in terms of a QUBO Hamiltonian:

$$O_{\text{QUBO}}(s_i) = \sum_{i,j=1}^{N_{\text{input}}} Q_{ij} s_i s_j, \tag{10}$$

where $s_i$ is the binary $\{0,1\}$ for each jet constituent to define which jet it is assigned to, $Q_{ij}$ is the QUBO matrix, which stores the distance between the $i$-th and $j$-th constituents, and $N_{\text{input}}$ is the number of inputs, namely the jet constituents. A QUBO Hamiltonian can be converted to an Ising Hamiltonian (Eq. (1)) by:

$$x_i \longleftrightarrow 2s_i - 1, \tag{11}$$

$$J_{ij} \longleftrightarrow \frac{Q_{ij}}{2}, \tag{12}$$

$$h_i \longleftrightarrow \frac{\sum_j Q_{ij}}{2}. \tag{13}$$

The thrust and quantum angle-based algorithms were considered in the previous studies [26–28]. In this study, the quantum angle-based algorithm is considered a benchmark and its QUBO matrix is defined as:

$$Q_{ij} = -\frac{\vec{p}_i \cdot \vec{p}_j}{2(|\vec{p}_i| \cdot |\vec{p}_j|)}, \tag{14}$$

where $\vec{p}_i$ is the momentum of the $i$-th jet constituent. It is compared to the QUBO matrix based on the $ee$-$k_t$ distance [50]:

$$Q_{ij} = 2\min(E_i^2, E_j^2)(1 - \cos\theta_{ij}), \tag{15}$$

where $E_i$ is the energy of the $i$-th jet constituent, and $\theta_{ij}$ is the angle between the $i$-th and $j$-th jet constituents. The $ee$-$k_t$ or the so-called Durham algorithm [50] is the standard jet-finding algorithm adopted in recent electron-positron colliders and is described in Section 2.3.

It is worth noting that the above formalism can only handle dijet clustering, as is evident from the binary implementation. In order to expand the method to multijet problems, the QUBO can be generalized to [26,28]:

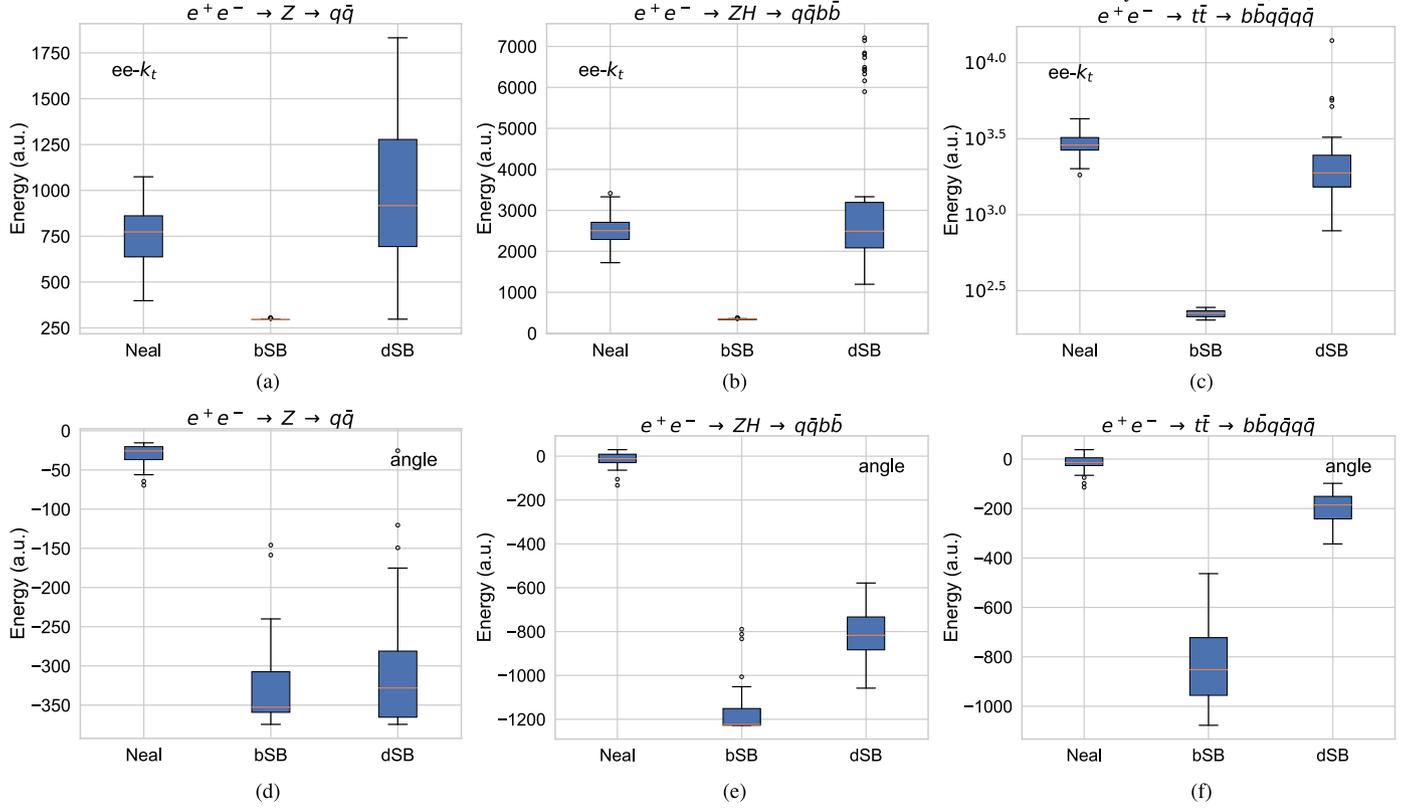

**Fig. 1.** Minimum Ising energies by the three QAIAs for a $Z$ boson (a, d), $ZH$ (b, e), and $t\bar{t}$ events (c, f) with the $ee$-$k_t$-based (a, b, c) or angle-based (d, e, f) distance considered in the QUBO formulation.

$$O_{\text{QUBO}}^{\text{multijet}}(s_i^{(n)}) = \sum_{n=1}^{n_{\text{jet}}} \sum_{i,j=1}^{N_{\text{input}}} Q_{ij} s_i^{(n)} s_j^{(n)} + \lambda \sum_{i=1}^{N_{\text{input}}} \left(1 - \sum_{n=1}^{n_{\text{jet}}} s_i^{(n)}\right)^2, \quad (16)$$

where $n$ considers the jet multiplicity and the binary $s_i^{(n)}$ is defined for each jet. The second term is introduced as the constraint to ensure that each jet constituent is assigned to a jet only once. The coefficient of this penalty $\lambda$ must be large enough: $\lambda > N_{\text{input}} \max_{ij} Q_{ij}$ [26]. In this study, the jet multiplicity is fixed to a specific value, as is the case for the exclusive jet finding pursued at the electron-positron colliders.

### 2.3. Benchmark algorithm

The Durham, or $ee$-$k_t$ algorithm, is adopted as our benchmark since it is the standard jet finder at the electron-positron colliders. It is implemented in the FASTJET software package [7]. The algorithm computes the distance $d_{ij}$ between every pair of inputs $i$ and $j$:

$$d_{ij} = 2\min(E_i^2, E_j^2)(1 - \cos\theta_{ij}). \quad (17)$$

It loops over iteratively, finds the smallest $d_{ij}$, and recombines the two inputs into a single "particle." The exclusive mode, which we consider in this study, terminates when the iteration reaches a fixed number of jets manually defined by the user. For the inclusive mode, usually adopted at hadron colliders, the user defines a scale $d_{\text{cut}}$, and the iteration stops when the minimum $d_{ij}$ exceeds the threshold. This study does not consider the latter since we only analyze the electron-positron collision events.

### 3. Dataset and event selection

We generate Monte Carlo (MC) simulated datasets for three physics processes, $e^+e^- \to Z \to q\bar{q}$, $e^+e^- \to ZH \to q\bar{q}b\bar{b}$ and $e^+e^- \to t\bar{t} \to bq\bar{q}'\bar{b}\bar{q}''q'''$ at the center-of-mass energy of 91 GeV, 240 GeV and 350 GeV respectively, with MADGRAPH_AMC@NLO [51] for the matrix element calculation, PYTHIA8 [52] (v8.2, GPL-2) for the parton showering and hadronization, and DELPHES [53] (v3.4.2, GPL-3) using the fourth detector concept [54] from the Circular Electron Positron Collider (CEPC) for the fast simulation of detector effects. The particle flow (EFlow) objects are considered as the jet reconstruction inputs. The three scenarios of the center-of-mass energy follow the CEPC proposal and are adopted here to evaluate the reconstruction performance for various jet multiplicities.

We only select events where all the jets are within the detector acceptance, namely $|\cos\theta| < 0.9$. Furthermore, the separations of two jets with the lowest transverse momenta in the events ($n$-th and $m$-th jets):

$$\sqrt{d_{nm}^{(\text{jet})}} = \sqrt{2\min(E_n^{(\text{jet})2}, E_m^{(\text{jet})2})(1 - \cos\theta_{nm})}, \quad (18)$$

are required to be larger than 20 GeV. This selection significantly suppresses background with QCD (gluon) radiation with a minimal impact on the signal [55] and is adopted as a baseline pre-selection in this study. In the $ZH$ and $t\bar{t}$ events, we require them to have exactly two $b$-tagged jets. To simplify the analysis mentioned in Section 4, only the $ZH$ events with the $Z$ bosons decaying to non-$b$ quarks are generated in the above MC simulation sample.

### 4. Results

In order to pursue jet clustering, QUBO Hamiltonians (Eq. (16)) are defined on an event-by-event basis. The jet multiplicity $n_{\text{jet}}$ is set to 2, 4, and 6 for the $Z$ boson, $ZH$, and $t\bar{t}$ production events. The binary $s_i^{(n)}$ for each jet constituent from the predicted ground state tells us whether the jet constituent is assigned to the $n$-th jet. Thus, the precision of the ground state prediction is the key to reconstructing the jets successfully.

First, the predicted minimum energy for Eq. (16) by the three QA-IAs are presented in Fig. 1 for a specific event from the three physics processes and two QUBO matrix definitions (Eqs. (14) and (15)) as examples. For all cases, bSB outperforms dSB and D-Wave Neal, i.e., in terms of the mean predicted energy and stability against multiple measurements. The trend becomes increasingly apparent as the physics process becomes more complicated, leading to an order-of-magnitude improvement in the minimum energy prediction for the $t\bar{t}$ process using the $ee$-$k_t$ distance. dSB, however, fails to predict minimum energy, and its performance is often comparable to that of D-Wave Neal. It is worth noting that the QUBO matrices for jet reconstruction are generally fully connected matrices, unlike the sparse QUBOs defined in track reconstruction [17–24]. QA is known for degraded performance in such a fully connected case [27], and remarkably, bSB can still provide quasi-optimal solutions.

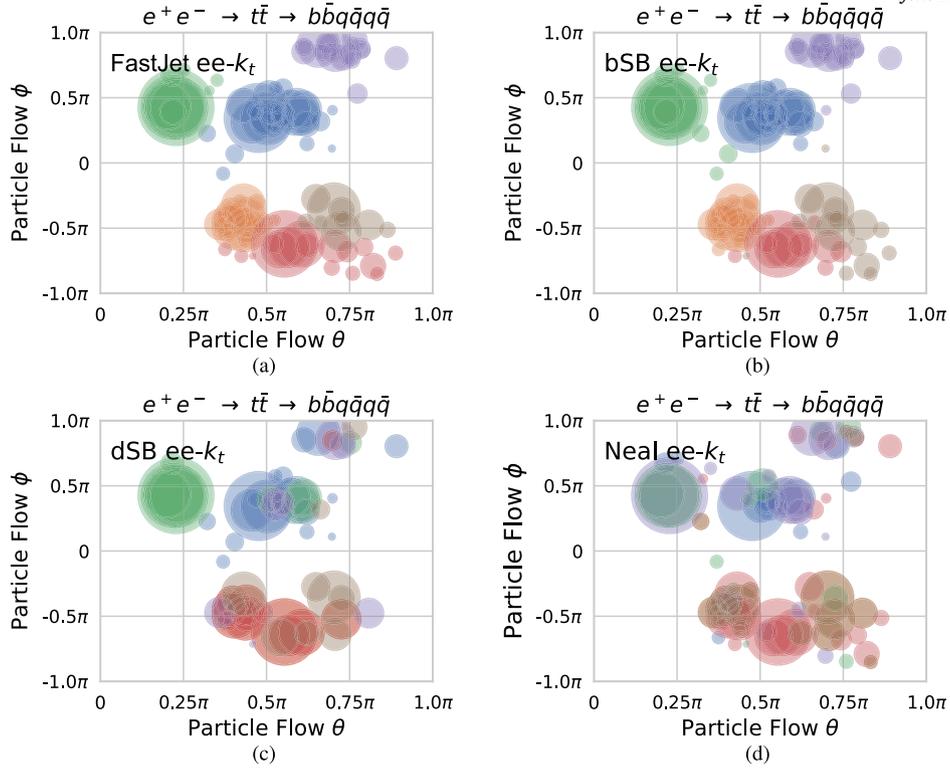

**Fig. 2.** Event displays from a $t\bar{t}$ event with jets reconstructed by (a) the $ee$-$k_t$ algorithm implemented in FASTJET or with the $ee$-$k_t$-distanced QUBO approach using (b) bSB, (c) dSB or (d) D-Wave Neal. Each circle represents a particle flow candidate with its size proportional to the energy. Each color corresponds to an individual jet.

In order to evaluate the performance of jet clustering algorithms, the most naive approach would be to compare with the true assignment of constituents to a jet corresponding to the original parton. However, as also stated in Ref. [31], it is impossible to define such an assignment in a reasonable manner. The fundamental issue is that the final state hadrons are occasionally hadronized from quarks originating from different initial partons. The situation will be even more complicated and fundamentally ambiguous when high-order calculations come into play. Thus, it is impossible to define the "true" particle-parton association in a meaningful way.

We adopt a commonly taken approach [28,31] to define a jet-constituent matching "efficiency" per jet by comparing to a corresponding classical algorithm, $ee$-$k_t$ implemented in FASTJET, and quantify the percentage of the jet constituents clustered in the same way:

$$\epsilon_{\text{jet}} = \frac{\text{\# of constituents clustered the same as FASTJET}}{\text{\# of constituents from FASTJET } ee\text{-}k_t}. \quad (19)$$

In the actual computation, the QAIAs return outputs of $n_{\text{jet}}$ arrays of jet constituent indices, where each array represents a jet. The jets reconstructed by FASTJET are recorded in the same manner to compute the efficiency. The number of overlapping array components between FASTJET and the QAIAs is used for the numerator in Eq. (19).

Fig. 2 shows event displays from a $t\bar{t}$ event using the benchmark $ee$-$k_t$ algorithm implemented in FASTJET and three QAIAs using the $ee$-$k_t$-distanced QUBO. The jet constituents are represented as circles, their size proportional to their energy. The colors represent the individual jets. Event displays with the angle-based QUBO, as well as for the other two physics processes, are presented in Appendix B. It is clearly seen that only bSB with the $ee$-$k_t$-distanced QUBO can successfully reconstruct jets for all three physics processes. FASTJET and bSB occasionally assign low-energy jet constituent outliers to different jets. Their impact is evaluated later in terms of the invariant mass resolution.

Fig. 3 shows the jet efficiencies (Eq. (19)) evaluated for the three physics processes using the three QAIAs with two types of QUBOs: $ee$-$k_t$-based (Eq. (15)) and angle-based (Eq. (14)). The figures show several important findings. First, the $ee$-$k_t$-based approach outperforms the angle-based counterpart for all the physics processes, which is partially expected due to the consistent definition of the distance in the $ee$-$k_t$-based QUBO. More importantly, the angle-based QUBO stops functioning for jet multiplicities beyond two; the jet efficiencies largely degrade in the $ZH$ and $t\bar{t}$ events, and even some jets have zero efficiency, meaning that they fail to be reconstructed despite the requirement of exclusive jet reconstruction. Most importantly, only bSB remains to succeed in jet reconstruction for all three physics processes. With dSB and D-Wave Neal, the constituents are often chaotically and unreasonably assigned to a jet, as can be seen in Figs. 2, B.6, B.7, and B.8. As stated above, fully-connected QUBOs are notoriously challenging to predict the minimum energy. bSB significantly outperforms dSB and Neal in this regard and demonstrates itself as a promising QUBO problem solver for handling multijet reconstruction.

As the benchmark $ee$-$k_t$ algorithm implemented in FASTJET does not necessarily provide the exactly "correct" answer, the jet efficiency alone does not give us an entirely decisive picture of the overall jet reconstruction performance. To evaluate the impact on the actual physics analysis, the invariant masses of the $Z$, Higgs bosons, and top quarks are presented in Fig. 4. In the $ZH$ events, the two $b$-tagged jets are assumed to originate from the Higgs bosons and are used in the mass reconstruction. The top-quark mass reconstruction is pursued in two steps, as was done in Ref. [56]: the two $b$-jets are assumed to originate from the $b$-quarks, and the light-flavor jet pairs with the least deviation from the $W$-boson mass $m_W$ are selected from the three possible permutations:

$$|m_{ij} - m_W| + |m_{kl} - m_W|, \tag{20}$$

where $i$, $j$, $k$, $l$ are the jet indices. Then, one of the two possible combinations of these light-flavor jet pairs and the $b$-jets more compatible with the top-quark mass are adopted. We did not consider the $\chi^2$ method [57], kinematic likelihood [58], or state-of-the-art machine learning methods [59–64] usually applied in hadron colliders, as the event topology and our assumption are simpler; the jet multiplicity is fixed to six, and the $b$-jets are assumed to come from the $b$-quarks.

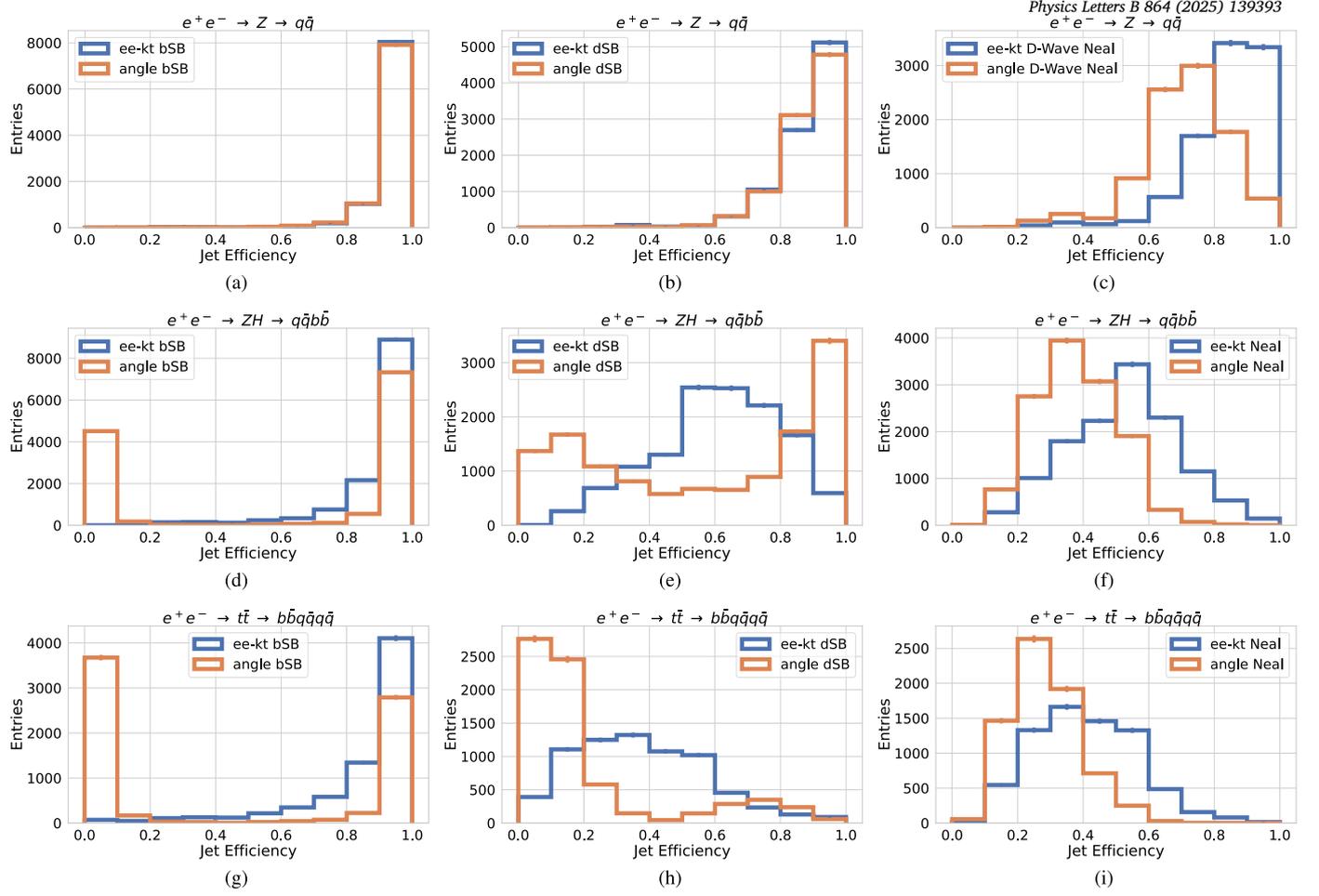

**Fig. 3.** Constituent-matching efficiencies to FASTJET for jets reconstructed by bSB (a, d, g), dSB (b, e, h), and D-Wave Neal (c, f, i) using $ee$-$k_t$ or angle-based QUBOs.

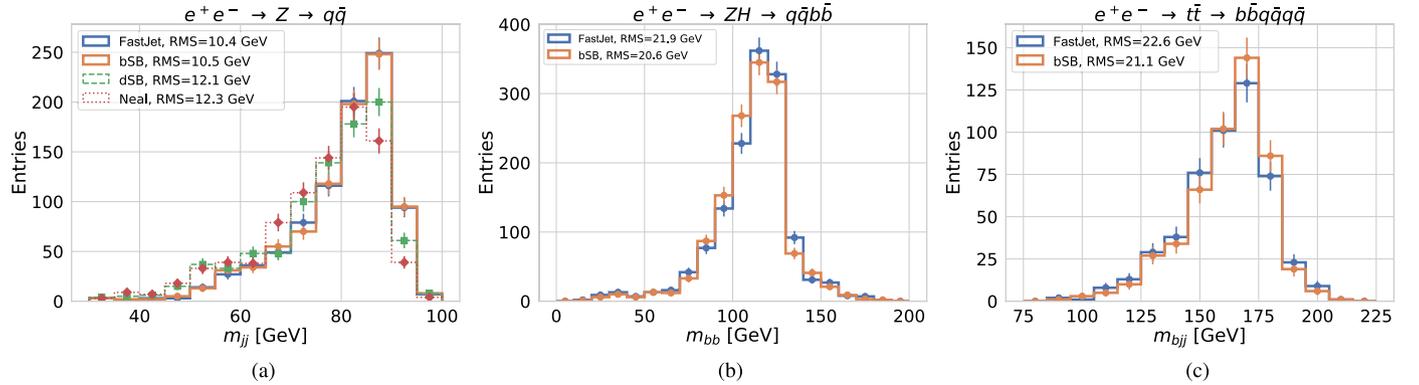

**Fig. 4.** Invariant masses of $Z$ bosons (a), Higgs bosons (b), and top quarks (c) with jets reconstructed by FASTJET and QAIAs. Only FASTJET and bSB are shown in (b, c), as dSB and Neal provide largely degraded performance and even fail to reconstruct some jets in the events.

FASTJET and bSB provide comparable performance in the $Z$ boson events, but dSB and D-Wave Neal already show visibly degraded mass resolution. In the $ZH$ and $t\bar{t}$ events, dSB and D-Wave Neal fail to reconstruct the jets reasonably as described above and are not shown in the figure. For these high-jet-multiplicity events (Figs. 4b and 4c), bSB provides slightly better mass resolution than the baseline FASTJET $ee$-$k_t$. The results indicate that the global QUBO jet reconstruction using bSB may provide more precise clustering.

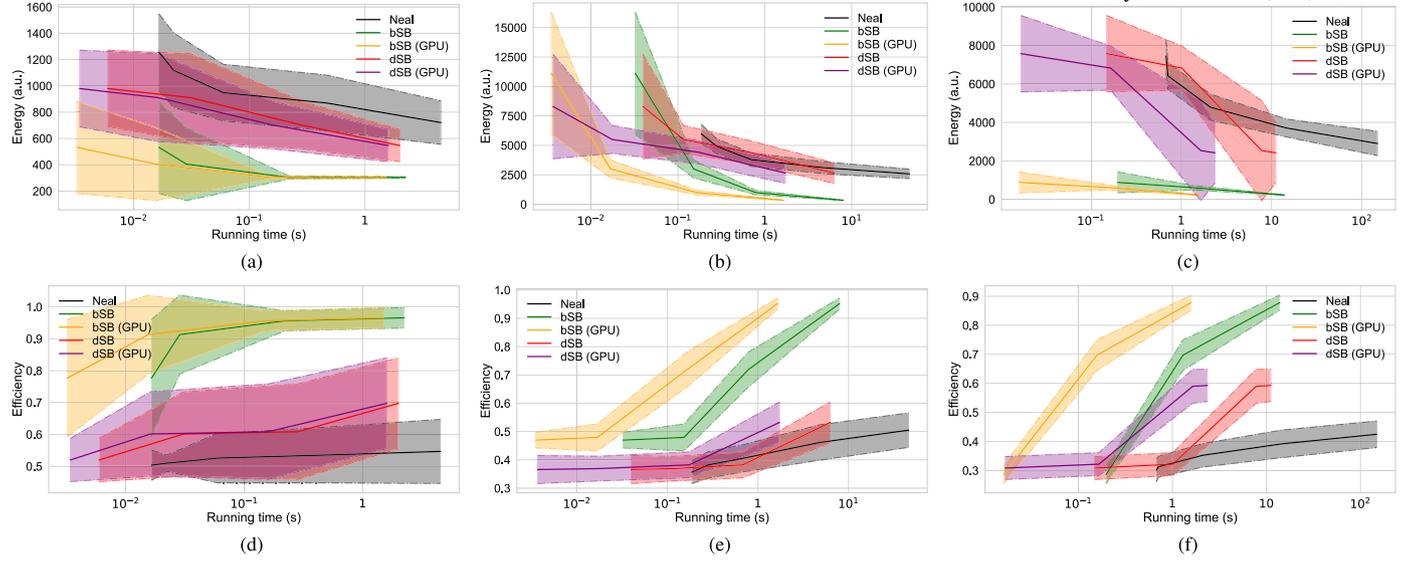

**Fig. 5.** Evolution of Ising energies (a, b, c) and jet efficiencies (d, e, f) against the running time from the three QAIAs presented for (a, d) $Z$ boson, (b, e) $ZH$, and (c, f) $t\bar{t}$ events. The $ee\text{-}k_t$ distance is adopted in the QUBO formulation. The solid lines are the average of multiple shots (100 for $Z$ and 50 for $ZH$ and $t\bar{t}$), and the envelopes represent the standard deviation from the mean.

**Table 1**
Mean running time for three physics processes and two target constituent-matching efficiencies using bSB with a CPU or GPU. The QUBO sizes correspond to the number of spins considered in the QAIAs and are equivalent to the number of qubits required when pursued on quantum computing hardware. dSB and D-Wave Neal do not reach the target efficiency even after a long duration, thus are not presented.

| Data Information | | | Mean running time [s] | |
| --- | --- | --- | --- | --- |
| Event | QUBO | Target eff. | bSB | bSB (GPU) |
| $Z$ | 64 | >0.8 | 0.02 | 0.006 |
| | | >0.9 | 0.03 | 0.02 |
| $ZH$ | 248 | >0.8 | 3.27 | 0.68 |
| | | >0.9 | 6.35 | 1.32 |
| $t\bar{t}$ | 444 | >0.8 | 8.43 | 0.97 |
| | | >0.9 | 15.36 | 1.76 |

The execution time for each QAIA algorithm is evaluated on an AMD Ryzen 7 6800HS Creator Edition CPU and an NVIDIA A100 GPU. Fig. 5 presents the evolution of Ising energies and jet efficiencies evaluated for the three QAIAs. We control the total runtime of the algorithms by setting different numbers of iterations. Only one CPU or GPU is used for a fair comparison with D-Wave Neal. The average of the 100 (50) shots for the $Z$ boson ($ZH$ and $t\bar{t}$) events and the envelopes defined from the standard deviation from the mean are shown in the figure. Table 1 shows the mean running times for the three physics processes extracted from Figs. 5d, 5e, and 5f. The QUBO sizes correspond to the number of qubits required for quantum hardware. bSB performs the best and rapidly converges to optimal values. In contrast, dSB and D-Wave Neal fail to reach reasonable jet efficiencies and Ising energies for all three physics processes regardless of the running time.

## 5. Conclusion and outlook

In our study, jet clustering is formulated as a QUBO problem. Three QAIAs are adopted to pursue global reconstruction, a new approach compared to the traditional iterative reconstruction implemented in FASTJET. The distance defined in the QUBO design significantly impacts the reconstruction performance, particularly when the jet multiplicity is beyond two. The angle-based approach only provides reasonable performance for dijet events, and alternative distances such as $ee\text{-}k_t$ are mandatory for higher jet multiplicities. For such multijet cases, it becomes exceedingly difficult even to approximately search for the "quasi"-ground state, ending up in a local minimum, which has an order-of-magnitude higher energy than the optimal states (Figs. 1b and 1c). Because of such a challenge, previous studies saw degraded performance in multijet reconstruction. However, this study shows that a powerful QUBO problem solver, bSB among the three QAIAs, can find optimal solutions. It demonstrates its outstanding capability to solve combinatorial optimization problems despite the high QUBO connectivities. It is promising that this global jet reconstruction with bSB can improve the invariant mass reconstruction in multijet events.

As a "quantum-inspired" algorithm, bSB runs on classical computers and is suitable for parallel processing and using cutting-edge computing resources such as GPUs and FPGAs. It is important to note that we can flexibly balance the speed and reconstruction precision in QAIAs. Namely, if precise energy resolution is not required for low-level triggers, for example, we can run reconstruction much faster. Thus, with the applicability to run on FPGAs, bSB may particularly provide an important option to be considered for triggers during the $Z$-pole data taking at the CEPC, an "exabyte"-level data taking comparable to the intensive HL-LHC conditions. The running time tends to be longer for high jet multiplicity events, and further investigations on the speed-up are ongoing.

Quantum annealing has been attempted for multijet reconstruction; however, using multiple qubits to implement one-hot encoding is challenging and prone to errors [28]. Furthermore, annealing time tends to have a long duration as estimated for D-Wave 2000Q with 6-way connectivity by a simulator with simplified jet datasets; it is approximately two orders of magnitude slower than bSB (Appendix A.1). It is to be seen with next-generation annealers such as D-Wave Advantage and Advantage2 with the increased qubit connectivity (15 and 20 ways respectively) whether we can address multijet reconstruction problems. In quantum gate machines, variational quantum algorithms such as Quantum Approximate Optimization Algorithm (QAOA) [65] have been actively investigated for optimization problems and are becoming competitors to quantum annealing [66]. However, QAOA generally lacks theoretical guarantees for quantum advantage and often suffers from Barren plateaus [67], especially when the problem sizes are large. Thus, QAOA often leads to degraded performance even with noise-less quan-tum simulators. Appendix A.2 shows the evaluation with the simplified jet datasets. Further investigations are required to understand whether we can successfully scale QAOA to larger problems with improved precision. Alternatively, qudits, multi-level computational units replacing the conventional 2-level qubits, may provide another approach to multijet reconstruction beyond the current binary optimization formulation [26].

Lastly, this study concentrates on $e^+e^-$ collider conditions and exclusive jet reconstruction, but an extension to inclusive jet reconstruction at hadron colliders would also be of interest and is left for future studies.

**Declaration of competing interest**

The authors declare that they have no known competing financial interests or personal relationships that could have appeared to influence the work reported in this paper.

**Acknowledgements**

HO would like to thank Gang Li, Shudong Wang, and Xu Gao for their feedback regarding the CEPC 4th detector concept in Delphes. HO is supported by the National Natural Science Foundation of China (NSFC) under Grant No. 12075060 and the NSFC Basic Science Center Program for Joint Research on High Energy Frontier Particle Physics under Grant No. 12188102.

**Appendix A. Prospects for quantum annealing and gate hardware**

We generate two simplified $Z$ boson datasets by storing only a fraction of the particle flow candidates with the highest transverse momenta, slimming it down to match 12 qubits from the original 68- and 90-qubit sizes. We use these simplified datasets for both quantum annealing and gate simulation studies described below.

*A.1. Simulation of quantum annealing hardware*

QuantumAnnealing.jl [68] is a Julia package that models analog quantum computer behavior, including D-Wave, on classical hardware. It simulates the time evolution of the Transverse Field Ising Model relying on the adiabatic theorem, with the Hamiltonian $H$ given by:

$$H(s) = A(s)H_{\text{initial}} + B(s)H_{\text{target}}, \tag{A.1}$$

where $s$ is the normalized time parameter, and $H_{\text{initial}}$ is the initial Hamiltonian of the system taken to be a transverse field, typically defined as:

$$H_{\text{initial}} = \sum_i \sigma_i^x. \tag{A.2}$$

$\sigma_i^x$ is the Pauli $X$ matrix operating on the $i$-th qubit. $H_{\text{target}}$ is the unknown ground state of the Ising Hamiltonian, representing the target optimization function:

$$H_{\text{target}} = \sum_{ij} J_{ij}\sigma_i^z\sigma_j^z + \sum_i h_i\sigma_i^z, \tag{A.3}$$

where $J_{ij}$ is the coupling strength between the $i$-th and $j$-th qubits, $\sigma_i^z$ is the Pauli $Z$ matrix operating on the $i$-th qubit, and $h_i$ is the strength of the external longitudinal field applied to the $i$-th qubit. $A(s)$ and $B(s)$ are the annealing schedules of the Hamiltonian. QuantumAnnealing.jl supports the D-Wave device-specific schedule (AS_DW_QUADRATIC), which directly mimics the physical behavior of the D-Wave 2000Q machine with 6-way connectivity at Los Alamos National Laboratory (LANL), making the simulation closer to real hardware. QuantumAnnealing.jl uses the Magnus expansion method to solve the time-dependent Schrödinger equation [69]. This method preserves unitarity, ensuring numerical solutions do not violate quantum state normalization and providing high accuracy.

**Table A.2**
Time-to-solution for D-Wave 2000Q estimated by simulation, bSB, dSB, and QAOA on a quantum circuit simulator for two simplified $Z \to q\bar{q}$ events.

| Event | D-Wave [s] | bSB [s] | dSB [s] | QAOA [s] |
|---|---|---|---|---|
| 0 | 21.29 | 0.35 | 0.79 | $1.07 \times 10^3$ |
| 1 | 20.52 | 0.36 | 0.89 | $3.36 \times 10^3$ |

We estimate the annealing time for the two simplified datasets described above. As we simplify the datasets to the 12-qubit size by only taking the particle flow candidates with the highest momenta, there is no ambiguity in the clustering, and the true solution is known. Thus, we consider time to solution (TTS) [70], the total time required to find the solution with sufficiently high probability, conventionally 99%, as the runtime metric.[1] Despite the small size of the simplified datasets, the estimated annealing time is around 20 seconds for both events (Table A.2), two orders of magnitude slower than bSB pursued on an AMD EPYC 7773X CPU.

*A.2. Quantum approximate optimization algorithm on quantum gate simulator*

We employ MindSpore Quantum (MindQuantum) [46,47] to simulate quantum circuits on classical computers. Quantum Approximate Optimization Algorithm (QAOA) is a variational quantum algorithm inspired by the Trotterization of the quantum adiabatic algorithm. It consists of a quantum component that prepares a quantum state according to a set of variational parameters and a classical component that optimizes the variational parameters and feeds them back to the quantum part in a closed loop. For each distinct parameter $p$, equivalent to the circuit depth, five sets of variational initial values for different QAOA circuit parameters are randomly generated, followed by gradient-based training until the convergence. Running the quantum circuits with the trained variational parameters yields a specific quantum state. This quantum state is used to calculate the expected energy and efficiency, and extract the probability amplitude of the correct state. Our results demonstrate that as the circuit depth $p$ increases, the expected energy from QAOA decreases, improving the expected efficiency and success probability. Table A.2 shows TTS with $p=14$; it takes more than three orders of magnitude longer than bSB on the MindQuantum circuit simulator. In principle, TTS could improve when operated on real quantum computing hardware, but it is left for future studies.

**Appendix B. Event displays**

Figs. B.6-B.8 present event displays for the three physics processes reconstructed with the benchmark FASTJET $ee$-$k_t$ algorithm or the bSB, dSB and D-Wave Neal using the $ee$-$k_t$ or the angled-based QUBOs.

**Appendix C. Supplementary material**

Supplementary material related to this article can be found online at https://doi.org/10.1016/j.physletb.2025.139393.

**Data availability**

The authors are unable or have chosen not to specify which data has been used.

---

[1] This metric cannot be directly compared to the mean running time in Table 1. We adopt TTS here as the problem is simplified and the solution is known. TTS could be significantly longer than the mean running time when there is a sizable fluctuation against multiple measurements.

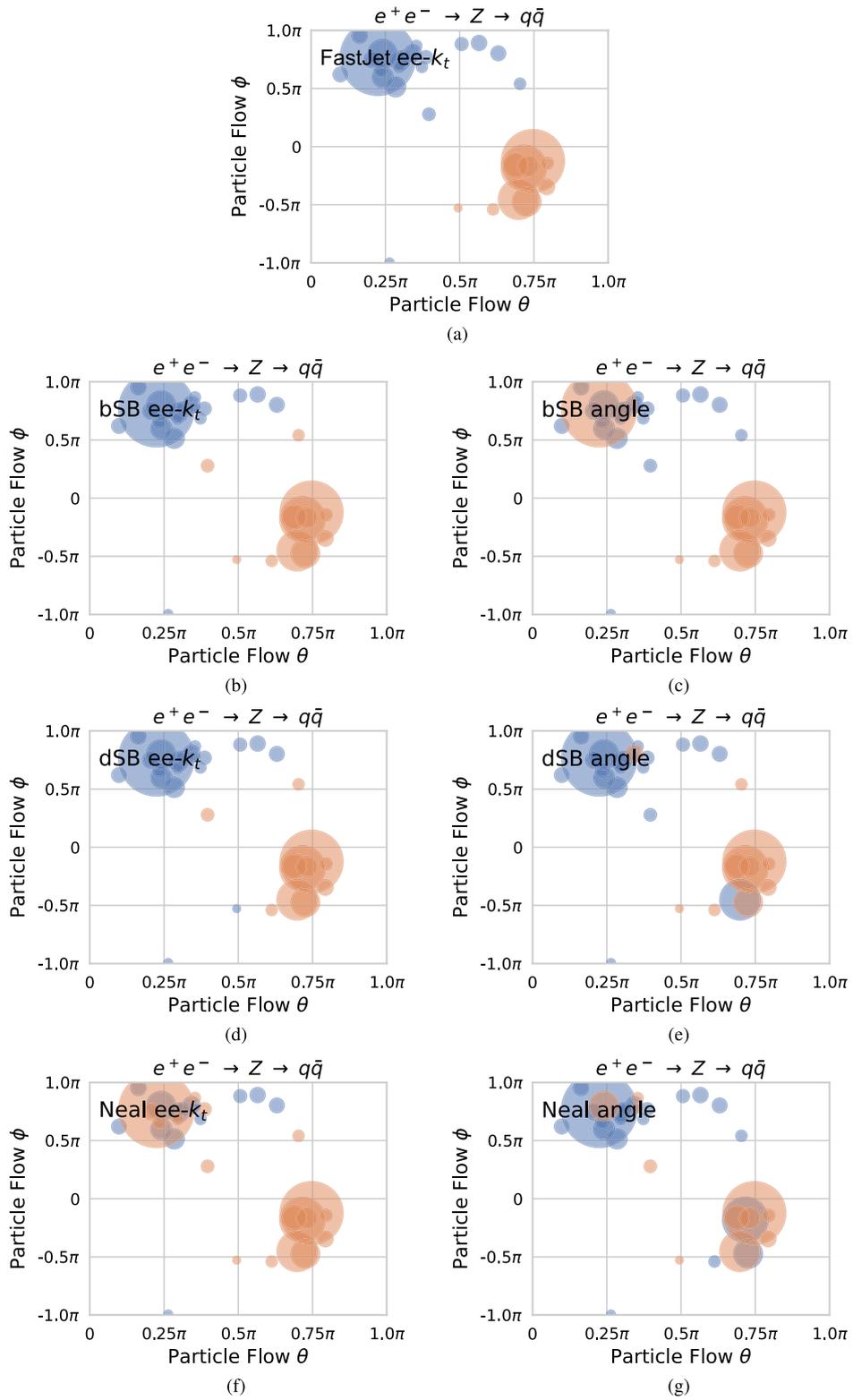

**Fig. B.6.** Event displays from a $Z$ event with jets reconstructed by (a) $ee$-$k_t$ algorithm implemented in FASTJET or with the $ee$-$k_t$- or angle-distanced QUBO approach using (b, c) bSB, (d, e) dSB or (f, g) D-Wave Neal. Each circle represents a particle flow candidate with its size proportional to the energy. The same color corresponds to the same jet.

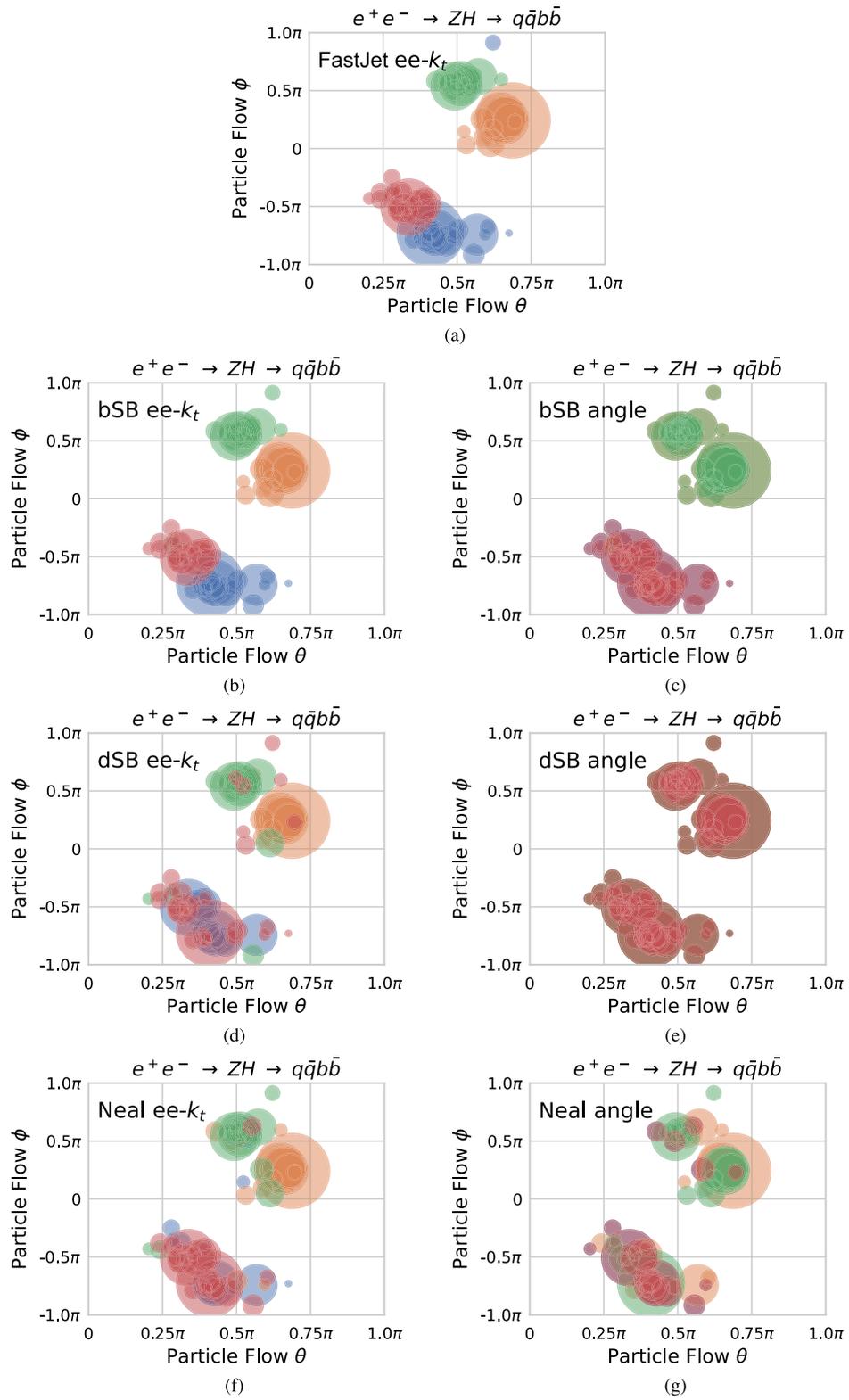

**Fig. B.7.** Event displays from a $ZH$ event with jets reconstructed by (a) $ee$-$k_t$ algorithm implemented in FASTJET or with the $ee$-$k_t$- or angle-distanced QUBO approach using (b, c) bSB, (d, e) dSB or (f, g) D-Wave Neal. Each circle represents a particle flow candidate with its size proportional to the energy. The same color corresponds to the same jet.

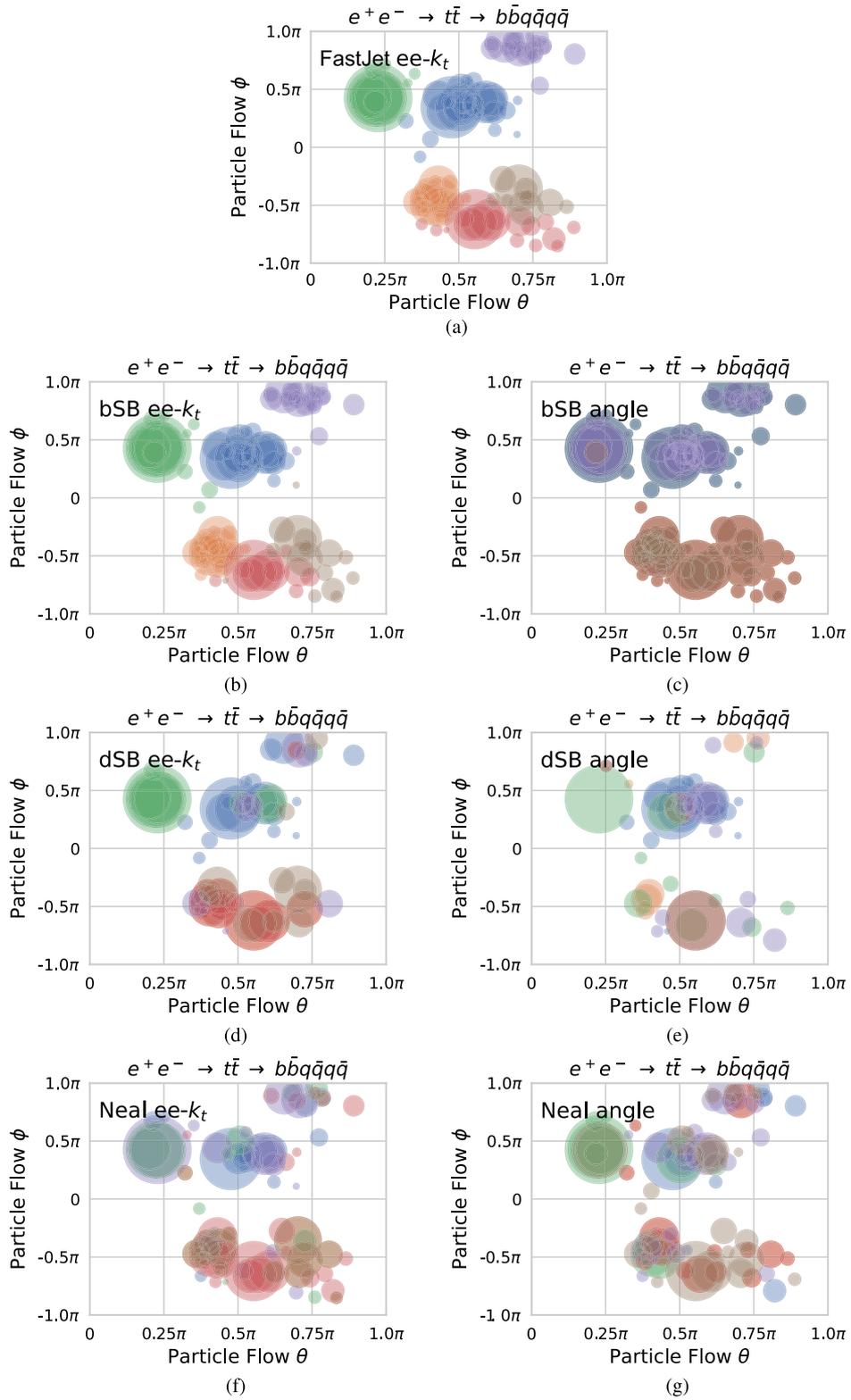

**Fig. B.8.** Event displays from a $t\bar{t}$ event with jets reconstructed by (a) the $ee$-$k_t$ algorithm implemented in FASTJET or with the $ee$-$k_t$- or angle-distanced QUBO approach using (b, c) bSB, (d, e) dSB or (f, g) D-Wave Neal. Each circle represents a particle flow candidate with its size proportional to the energy. The same color corresponds to the same jet.